\documentclass{emulateapj}
\usepackage{threeparttable}
\usepackage{natbib}
\usepackage[bookmarks=false, colorlinks=true, citecolor=blue,
  backref=true]{hyperref}              
\hypersetup{
  colorlinks,
  citecolor=blue,
  linkcolor=red,
  urlcolor=blue,
}

\DeclareRobustCommand{\ion}[2]{%
\relax\ifmmode
\ifx\testbx\f@series
{\mathbf{#1\,\mathsc{#2}}}\else
{\mathrm{#1\,\mathsc{#2}}}\fi
\else\textup{#1\,{\mdseries\textsc{#2}}}%
\fi}

\shorttitle{ }
\shortauthors{Cao et al.}

\begin{document}
\slugcomment{{\bf Accepted for publication in ApJ}}

\title{understanding the chemical evolution of blue Edge-on Low Surface 
Brightness Galaxies}
\author{
Tian-wen Cao\altaffilmark{1,2}, Hong Wu\altaffilmark{2}, Gaspar Galaz\altaffilmark{3}, 
Venu M. Kalari\altaffilmark{4}, Cheng Cheng\altaffilmark{2,5}, Zi-Jian Li\altaffilmark{2,5}, Jun-feng Wang\altaffilmark{1}
}
\altaffiltext{1}{Department of Astronomy, Xiamen University, 422 Siming South Road, Xiamen 361005, People$'$s Republic of China;\,jfwang@xmu.edu.cn}
\altaffiltext{2}{National Astronomical Observatories, Chinese Academy of Sciences, Beijing 100101, People$'$s Republic of China;\,hwu@bao.ac.cn}
\altaffiltext{3}{Instituto de Astrofisica, Pontificia Universidad  Cat\'olica de Chile,\,$\!$\,Av.$\!$Vicu\~na Mackenna$\!$\,4860,\,7820436 Macul,$\!$\,Santiago, Chile}
\altaffiltext{4}{Gemini Observatory/NSFs NOIRLab, Casilla 603, La Serena, Chile}
\altaffiltext{5}{Chinese Academy of Sciences South America Center for Astronomy, National Astronomical Observatories, Chinese Academy of Sciences, Beijing 100101, People$'$s Republic of China}

\begin{abstract}

We present a sample of 330 blue edge-on low surface brightness galaxies (ELSBGs).
To understand the chemical evolution of LSBGs, we derived the gas-phase abundance and the [$\alpha$/Fe] ratio.
Compared with star-forming galaxies, ELSBGs show a flatter trend in the mass-metallicity ($M_*-Z$) relation,
suggesting that the oxygen abundance enhancement is inefficient.
We focus on 77 ELSBGs with {\ion{H}{I}} data and found the closed-box model can not explain 
their gas fraction and metallicity relation, implying that infall and/or outflow is needed.
We derived the [$\alpha$/Fe] ratio of normal ELSBG ($<$ 10$^{9.5}$M$\odot$) and 
massive ELSBG ($>=$ 10$^{9.5}$M$\odot$) using single stellar population grids from MILES stellar library.
The mean [$\alpha$/Fe] ratios are 0.18 and 0.4 for normal ELSBG and massive ELSBG, respectively.
We discussed that the long time-scale of star-formation, and/or metal-rich gas outflow event caused 
by SNe\,Ia winds are likely responsible for the $\alpha$-enhancement of massive ELSBGs.

\end{abstract}

\keywords{galaxies: abundances - galaxies: evolution - galaxies: stellar populations}

\section{Introduction \label{intro}}
Low surface brightness galaxies (LSBGs) provide a unique insight into the faint end.
The definition of LSBG is that the central surface brightness ($\rm \mu_{0}$) is 
at least one magnitude darker than the night sky brightness in B-band, usually 22.0-23.0 
mag arcsec$^{-2}$ \citep{1997ARA&A..35..267I}.
LSBGs have been discovered both in galaxy clusters and field environments
\citep{2015ApJ...813...68G, 2016MNRAS.456.1607D, 2018ApJ...857..104G, 2020MNRAS.496.3182A},
and they present peculiar properties.
The star-formation rate of LSBGs is lower than main-sequence galaxies 
\citep{2009ApJ...696.1834W, 2018ApJS..235...18L, 2019ApJS..242...11L, 2022arXiv221104342G} and LSBGs are metal-poor 
\citep{2004MNRAS.355..887K, 2017ApJ...837..152D}. 
Some LSBGs are {\ion{H}{I}} rich \citep{2015AJ....149..199D, 2020ApJS..248...33H}. 
Nevertheless, many of them lack molecular gas \citep{2001ApJ...549L.191M, 2008ApJ...677L..13G, 2010A&A...523A..63D, 2017AJ....154..116C}.
The fraction of active galactic nuclei (AGN) in LSBGs is lower than that of high surface brightness galaxies \citep{2011ApJ...728...74G}.
The nuclear activity is weak in nearby LSBGs \citep{2020ApJ...898..106H}.
How LSBGs form and evolve remains an open question, 
due to their inherent faintness leading to observational challenges.
With the advent of large photometric and spectroscopic surveys, some of these issues can be overcome. 
It then becomes important and feasible to understand how LSBGs occur in diverse environments  
or have varying evolutionary paths as this informs cosmological and galaxy evolution models.

With expanded optical sky survey data releases,
there are enlarged LSBG samples \citep{2016MNRAS.463.2746W, 2018ApJ...857..104G, 2021ApJS..252...18T} 
that permit better statistics on their properties.
The deep and wide-field imaging surveys, such as the Dark Energy Survey (DES; \citealt{2015AJ....150..150F}), 
the Kilo-Degree Survey (KIDs; \citealt{2015A&A...582A..62D}), 
and the Hyper Suprime-Cam Subaru Strategic Program (HSC-SSP; \citealt{2018PASJ...70S...8A}) provide 
an unprecedented opportunity to detect LSBGs.

The spectroscopic surveys are still very deficient for LSBGs 
due to their faint and diffuse structure. 
Limited by spectra observations, the chemical evolution of LSBGs is not fully discussed 
\citep{1994ApJ...426..135M, 2001AJ....122.2318B, 2002AJ....124.1360G,
2004MNRAS.355..887K, 2010A&A...520A..69Z, 2015MNRAS.446.4291L, 2017ApJ...837..152D}. 
Investigating chemical evolution can help us reconstruct the star-formation histories of galaxies. 
The canonical closed-box model is a kind of secular evolution of chemical abundance \citep{1963ApJ...137..758S, 1972ApJ...173...25S}
which describes that the initial metal-free gas is permanently locked into
stars and there is neither an inflow nor outflow process.
The chemical evolution of those gas-rich and disk-dominant LSBGs is not 
along the closed-box pathway according to the relationship between 
the near-infrared (nIR) color and M$_{\rm HI}$/M$_{\rm baryonic}$ \citep{2002AJ....124.1360G}. 
\cite{2002AJ....124.1360G} speculated that those LSBGs are possibly experiencing the metal-poor gas
infall and dilute the metallicity of subsequent generations of stars. 
However, a sample of LSBGs from \cite{2004MNRAS.355..887K} shows their chemical evolution is consistent with 
the closed-box model by studying {\ion{H}{II}} region spectra. 
Gas-phase abundance studies help create the complicated chemical evolution models of LSBGs\citep{2017ApJ...837..152D}.
Besides gas-phase abundances, one can obtain recent star-formation histories through metallicity of the stellar population. 
Namely, the ratio of $\alpha$ elements, which are produced during Type II supernovae ({\ion{SNe}{II}}), 
to iron that mainly comes from Type Ia supernovae (SNe\,Ia). [$\alpha$/Fe] can trace the star-formation activities, 
since the timescale from the star formation to the explosion of {\ion{SNe}{II}} is about a few tens of million years,
which is much shorter than that of SNe\,Ia (0.1$\sim$1 Gyr) \citep{2011MNRAS.413L...1C}. 
The $\alpha$ elements can be enhanced by the short episode star-formation activity and
can be diluted when SNe\,Ia adds iron elements. 

The Sloan Digital Sky Survey (SDSS; \citealt{2000AJ....120.1579Y, 2002AJ....123..485S, 2003AJ....126.2081A, 2004AJ....128..502A})
provides a complete spectroscopic survey at r-band 17.7 mag \citep{2002AJ....124.1810S}.
We need a bright LSBGs sample to explore their spectroscopic properties by SDSS.
Edge-on galaxies have higher surface brightness than face-on galaxies. 
Furthermore, ELSBG can help us to understand the formation and evolution of LSBGs by dynamics 
without uncertainty in inclination.
So, an ELSBGs sample with spectra is necessary for comprehensive understanding LSBGs.
\cite{2020ApJS..248...33H} presented a {\ion{H}{I}}-rich edge-on low surface brightness galaxies (ELSBGs) sample 
but lacked spectra observations. 

In this paper, we focus on the chemical abundance properties of a sample of 330 blue ELSBGs.
We describe the sample selection in Section\,2. 
In Section\,3, we discuss the chemical abundance evolution and stellar population content of our ELSBGs. 
We summarize this work in Section 4. All magnitudes in this paper are AB magnitudes. 
We adopt a $\Lambda$CDM cosmology with $\rm \Omega_{m}$ = 0.3, 
$\rm \Omega_{\Lambda}$ = 0.7, $\rm H_0$ = 70\,km\,s$^{-1}$\,Mpc$^{-1}$.

\section{Sample selection \label{sample}}
\subsection{ Pre-selection}
\citet{2011ApJS..196...11S} presents a catalog of 1.12 million galaxies from SDSS DR7 \citep{2009ApJS..182..543A}. 
The catalog is complete at 17.7 mag in r-band with SDSS fiber spectra.
We select our initial sample using the structural parameters from Table\,2 of \citet{2011ApJS..196...11S}.
The free-n$\rm _b$ bulge + disk compositions model is applied in Table\,2.
The criteria are (i) the inclination of disks larger than 80 degrees to select edge-on galaxies
 and (ii) their n$\rm _b$ smaller than two to ensure disk dominant galaxies.
There are 6660 targets in our initial sample.

We apply the edge-on model to fit our initial sample with GALFIT \citep{2002AJ....124..266P} 
of SDSS g-band.
The initial parameters refer to the catalog of \citet{2011ApJS..196...11S} and the output parameters 
from GALFIT including major axis disk scale length (r$_s$), 
the perpendicular disk scale height (h$\rm _s$), and projected central surface brightness ($\rm \mu_{0-edge}$). 
The face-on center surface brightness ($\rm \mu_{0-face}$) can be estimated by
 $\rm \mu_{0-face}$ = $\rm \mu_{0-edge}$ $-$ 2.5$\rm \times$log$\rm _{10}$(h$\rm _s$/r$\rm _s$) \citep{2020ApJS..248...33H}. 
To avoid being dependent on the model, we measure the average surface brightness 
in 3$''$ radius ($\rm \mu_{3-edge}$) of the center region
and perform the aperture photometry by the Photutils package \citep{larry_bradley_2020_4044744}.
Since the flux of a face-on disk is the integration of the perpendicular disk flux.
We assumed that the perpendicular disk profile is $\Sigma$(z)= sech$^2$(h/h$\rm _s$) \citep{2005AAS...207.8705S}
to correct the projected edge-on flux in 3$''$ radius to the perpendicular disk flux. 
Then, we convert the corrected edge-on $\rm \mu_{3-edge}$ to face-on $\rm \mu_{3-face}$ by assuming
 an exponential disk profile of our targets and simplify this calculation in two-dimension. 
We select $\rm \mu_{3-face}$ fainter than 25 mag\,arcsec$\rm ^{-2}$ 
into the pre-selection sample and there are 1409 targets.

\begin{figure}
  \begin{center}
  \includegraphics[width=3.5in]{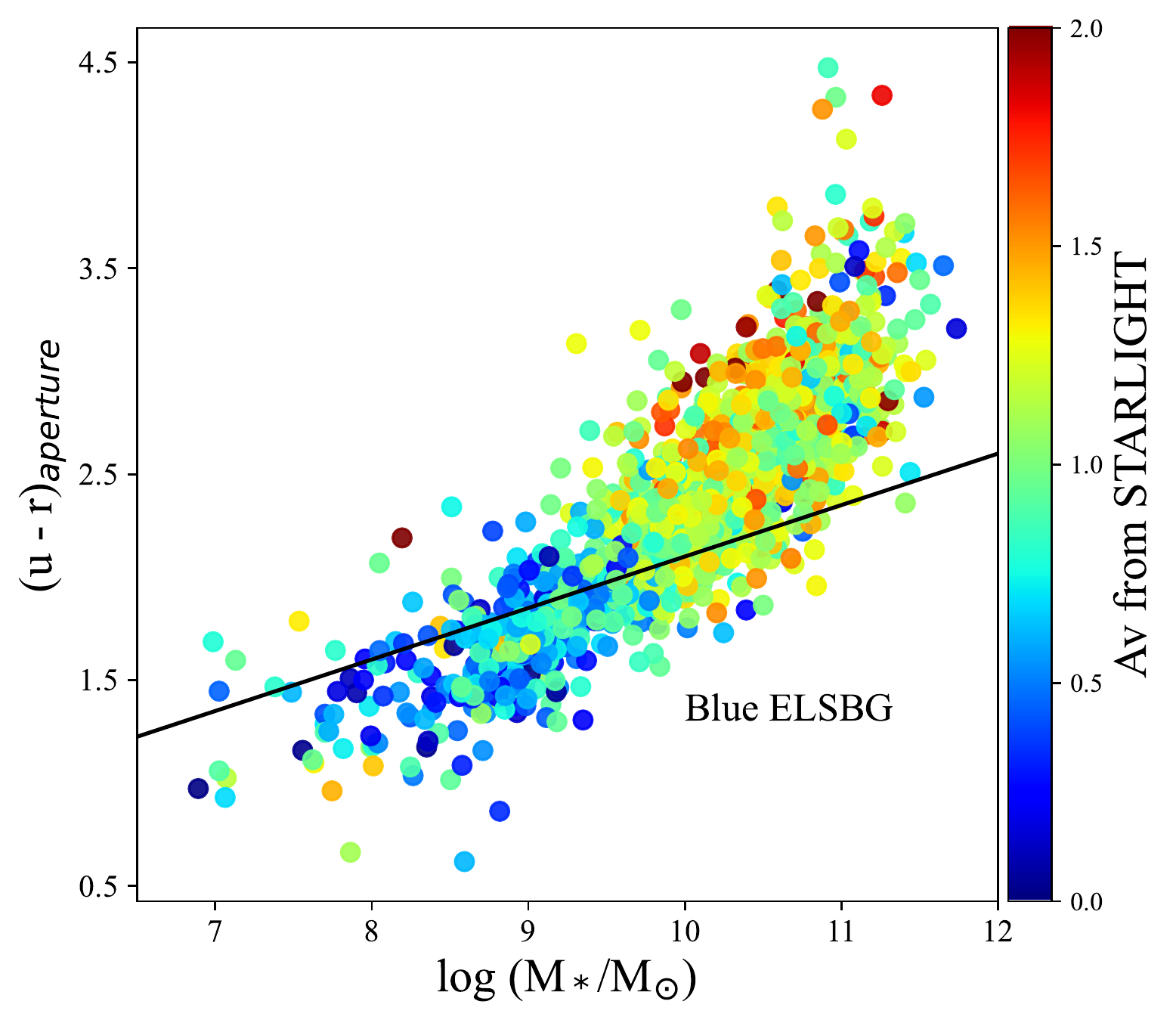}
  \end{center} 
\caption{The observed (u - r) color and mass relation (CMR) for our pre-selected sample.
The (u - r) color is from the center 3$''$ aperture region.
The black line (u -r) = -0.4 + 0.25$\times$log(M$_*$/M$\odot$) is from \cite{2021ApJ...908....4V} 
and there are 367 blue ELSBGs under the black line. 
The color code shows the continuum of dust extinction from STARLIGHT.
   \label{colormass}}
\end{figure}

\subsection{ Blue Edge-on Low Surface Brightness Galaxies}
\begin{figure}
  \begin{center}
  \includegraphics[width=3.2in]{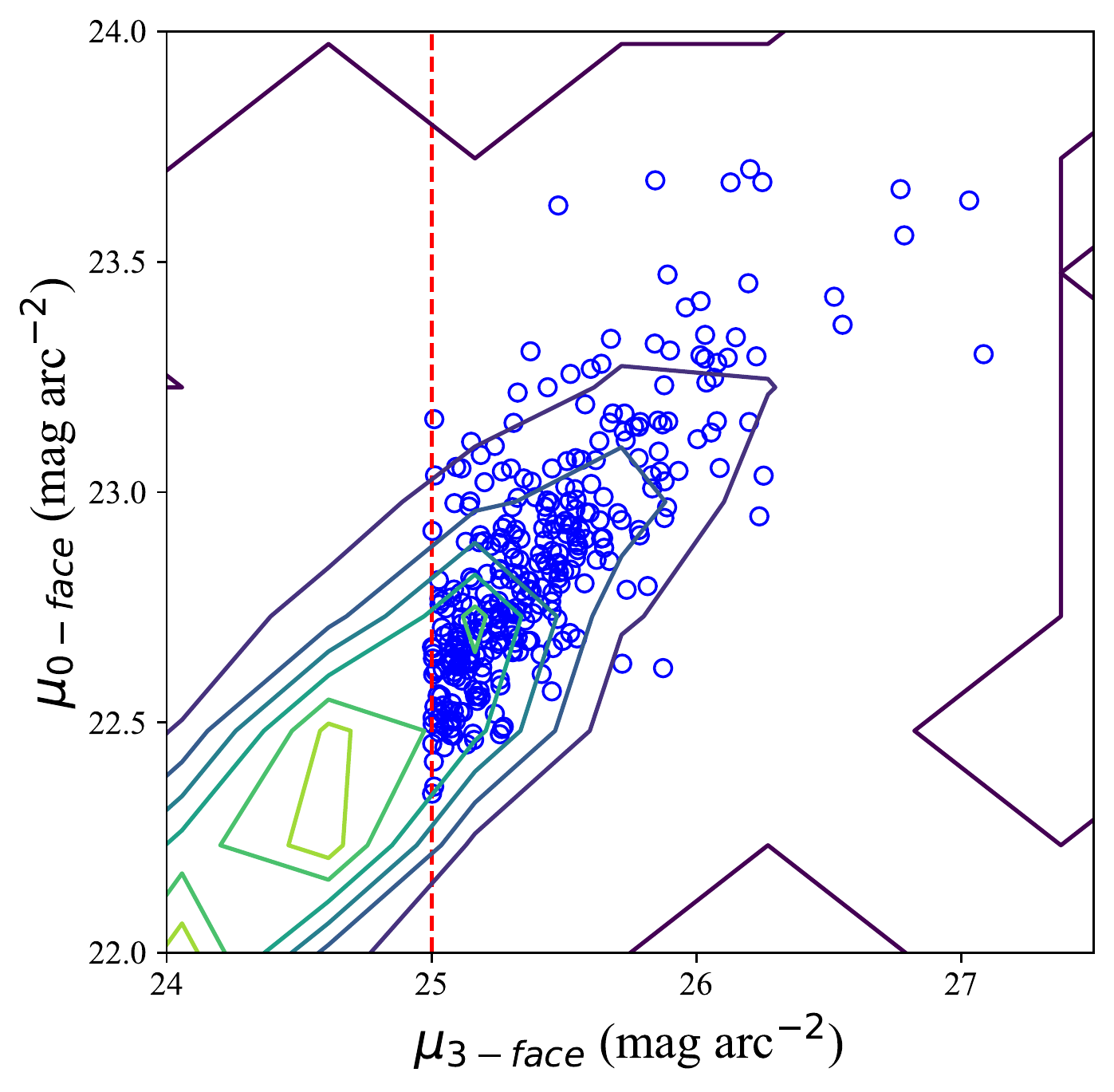}
  \end{center}
\caption{The $\rm \mu_{0-face}$ versus $\rm \mu_{3-face}$.
Contours represent the initial sample and open blue circles represent our final 330 blue
ELSBGs}. The dashed red line is $\rm \mu_{3-face}$ = 25 mag arcsec$^{-2}$.
           \label{selection}
\end{figure}

We select blue ELSBGs by the observed (u - r) color and stellar mass relation (CMR).
We measured the (u - r) color of 3$''$ aperture (consistency with SDSS fiber region) in the central region, 
and the stellar mass is from MPA-JHU catalog \citep{2003MNRAS.341...33K},
 which is calculated by the z-band mass-to-light (M/L) ratio based on multi-band SED fitting.
Figure\,\ref{colormass} shows the CMR of the pre-selection sample.  
With stellar mass increasing, the center (u - r) color shows an increasing trend. 
We apply the green valley upper boundary from \cite{2021ApJ...908....4V} to select blue ELSBGs 
and there are 367 blue ELSBGs.
Taking advantage of SDSS spectra, we apply STARLIGHT \citep{2005MNRAS.358..363C} 
software to check dust extinction.
We give the detailed setting of STARLIGHT in Section\,3.3. 
Galaxies with stellar masses larger than 10$^{10}$ M$_{\odot}$ 
usually experience heavier extinctions than other ELSBGs in Figure\,\ref{colormass}.
Most blue ELSBGs show their Av values smaller than 1.0 Mag.
We check their SDSS color images and exclude 37 targets that are not edge-on galaxies or polluted by a very bright star. 
There are 330 blue ELSBGs in our final sample shown in Figure\,\ref{selection}.
$\rm \mu_{0-face}$ of our sample ranges from 22.34 to 23.75 mag\,arcsec$\rm ^{-2}$. 
Since the extinction in Figure\,\ref{colormass} is cumulative dust extinction along the disk, 
the corresponding face-on center should have less extinction.
Therefore, dust extinction can not have a significant effect on the center surface brightness.

\begin{figure*}
  \begin{center}
  \includegraphics[width=7.0in]{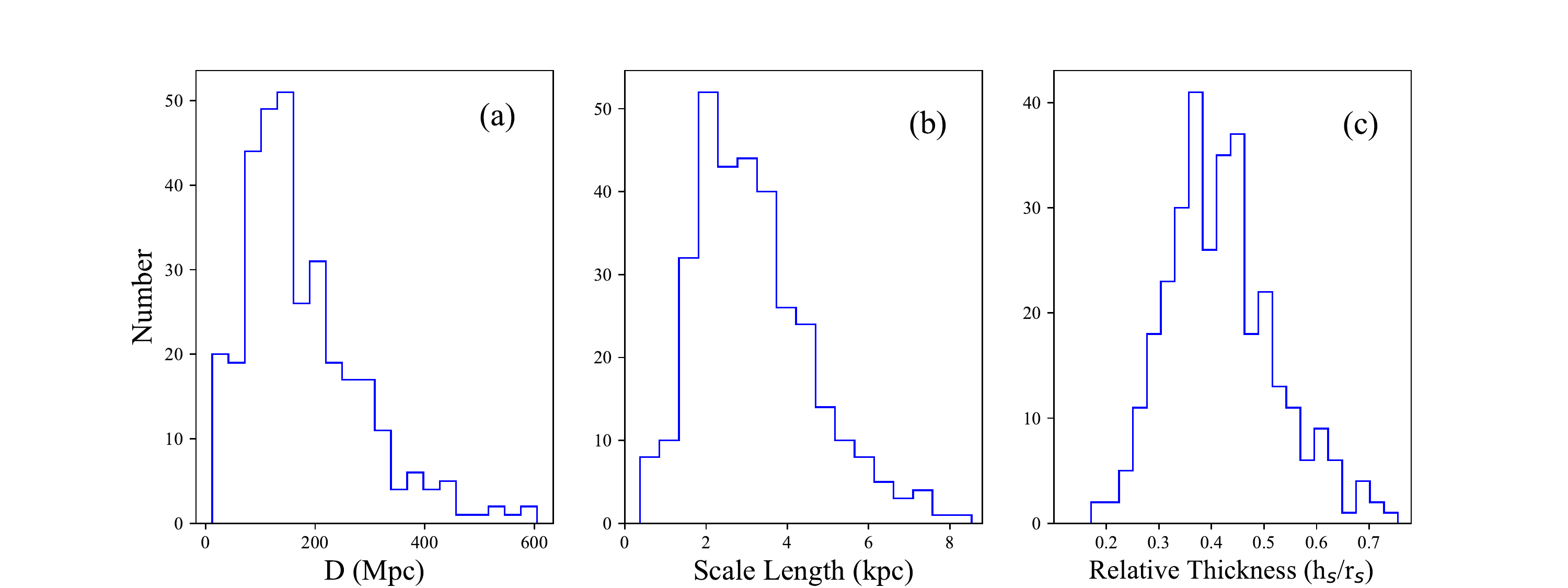}
  \end{center}
\caption{Histograms of relevant physical parameters of 330 ELSBGs.  
Panel (a): the distribution of distance in Mpc. The distance is simply estimated by $cz$/H$_0$. 
 Panel (b): the distribution of SDSS g-band scale length (r$\rm _s$) in kpc from GALFIT. 
Panel (c): the distribution of SDSS g-band relative thickness (h$\rm _s$/r$\rm _s$) from GALFIT.}
       \label{parameter}
\end{figure*}

The redshifts of 330 blue ELSBGs range from 0.0026 to 0.14. 
We show the distribution of their distance, physical scale length (r$\rm _s$), 
and relative thickness in Figure\,\ref{parameter}. 
The generally accepted scale length value of the Milky Way is about 2.3 kpc in R-band \citep{2007ApJ...662..322H}.
Most of our blue ELSBGs show extended disks in SDSS g-band compared with the Milky Way.

\section{Analysis and Discussion}
\subsection{($M_*-Z$) Relation}

Stellar mass and metallicity are two fundamental parameters 
of galaxies because they are related to the evolution of galaxies. 
The gas-phase metallicity can be derived from different nebular line diagnostics 
 (e.g. H$\alpha$, H$\beta$, [{\ion{N}{II}}], [{\ion{O}{III}}], \citealt{2004MNRAS.348L..59P, 2013A&A...559A.114M}).  
We derived the fiber metallicity 12+log(O/H) by [{\ion{N}{II}}]$\lambda$6583/H$\alpha$ (N2) index \citep{2004MNRAS.348L..59P}. 
Since [{\ion{N}{II}}] emission line is much weaker than H$\alpha$ emission line, 
we estimate S/N of [{\ion{N}{II}}] emission line to exclude targets with very weak emission lines.
We subtract the continuum by STARLIGHT fitting result.
The noise level is considered as the standard deviation of continuum-subtracted spectrum $\sigma$ (6500$\AA$-6650$\AA$).
We use the Gaussian model to fit the [{\ion{N}{II}}] emission line and set the line region of 3$\sigma_{line}$.
The S/N is defined as line flux over $\sqrt{N}*\sigma$, where N is the pixel number in the line region.
310 blue ELSBGs show S/N of [{\ion{N}{II}}] emission lines larger than three. 
We measure their [{\ion{N}{II}}] and H$\alpha$ fluxes based on their continuum-subtracted spectra and estimate 
their gas-phase metallicities.

\begin{figure}
  \begin{center}
  \includegraphics[width=3.5in]{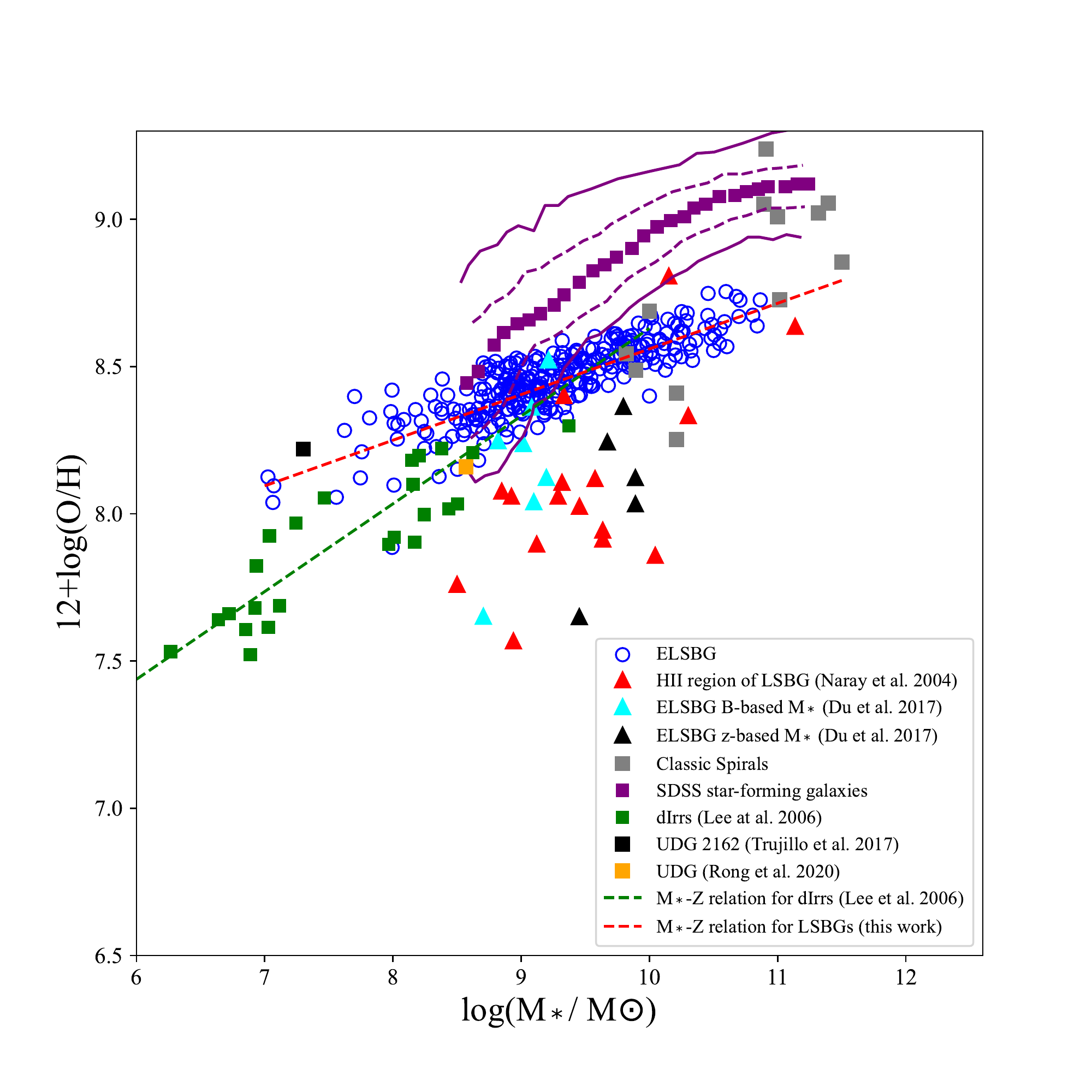}
  \end{center}
\caption{Stellar mass versus gas-phase metallicity. 
Our ELSBGs are shown as open blue circles and the blue line is the linear ($M_*-Z$) relation fit for our sample
(12 + log(O/H) = 7.01 + 0.155 $\times$ log(M$_*$/M$\odot$) with the dispersion of 0.068 dex).
Red triangles are LSBGs from \citet{2004MNRAS.355..887K}. We show ELSBGs from \citet{2017ApJ...837..152D}
shown as cyan and black triangles, respectively. Cyan and black triangles represent B-band and z-band based M$_*$.
The grey squares are normal massive spiral galaxies 
from \cite{1987ApJ...317...82G}, and the purple squares represent the medians of the 53,000 
SDSS star-forming galaxies from \citet{2004ApJ...613..898T} in bins of 0.1 dex in stellar mass. 
The purple dashed and solid lines are, respectively, contours 
that enclose 68\% and 95\% of the 53,000 star-forming galaxies.
The green squares represent the dwarf irregular galaxies (dIrrs) and the green line is the linear ($M_*-Z$) relation fit for dIrrs 
from \citet{2006ApJ...647..970L}. 
The ($M_*-Z$) relation for dIrrs is 12 + log(O/H) = 5.65 + 0.298 $\times$ log(M$_*$/M$\odot$).}
           \label{metal}
\end{figure}

Figure\,\ref{metal} shows stellar mass versus metallicity ($M_*-Z$) for different types of galaxies.
Spectra of our blue ELSBGs and SDSS star-forming galaxies were taken using fiber spectroscopy, 
wherein the aperture of the fiber covers mainly the central regions. 
We also show metallicities of the entire ELSBGs \citep{2017ApJ...837..152D} in Figure\,\ref{metal}.
The metallicity is derived from N2 index of \citet{2017ApJ...837..152D}. 
Since \cite{2017ApJ...837..152D} applied different method to estimate 
the stellar masses, we show both B-band based and z-band based M$_*$ in Figure\,\ref{metal}. 
The metallicity of \citet{2004MNRAS.355..887K} is measured from {\ion{H}{II}} regions, and the stellar mass
is based on B-band luminosity.
We note that the results of both \cite{2004MNRAS.355..887K} and \cite{2017ApJ...837..152D} are from long-slit spectra.
The metallicity of our sample is higher than that of entire ELSBGs and {\ion{H}{II}} regions.
\citet{2021RAA....21...76C} presented that HI-dominated LSBG has a similar H-band size (stellar structure) 
to other local galaxies with a more extended star-formation radius. 
The long slit spectra of ELSBGs in \cite{2017ApJ...837..152D} may be 
dominated by that outer skirt {\ion{H}{II}} region.
This may be the reason that the metallicity of \cite{2017ApJ...837..152D} is 
consistent with that of \citet{2004MNRAS.355..887K}. 
The extinction of our sample is not significant in Figure\,\ref{colormass}.
So, the fiber spectrum of our sample could better represent the abundance of ELSBG 
in the center region.

Galaxies usually first build their stellar structure 
in the center region. This means that chemical evolution should first happen in the center.
Our blue ELSBGs and SDSS star-forming galaxies sample cover a similar redshift (z $\sim$ 0.1) 
and stellar mass arrangements. We mainly compare the center region gas-phase metallicity between our blue ELSBGs 
and SDSS star-forming galaxies in Figure\,\ref{metal}. 
Our ELSBGs show lower metallicities than SDSS star-forming galaxies.
Metallicities of our ELSBGs range from 7.88 to 8.75 dex, with a mean value of 8.45 dex.
We obtain a linear ($M_*-Z$) relation for our blue ELSBGs: 12+log(O/H) = 7.01 + 0.155 $\times$ log(M$_*$/M$\odot$) 
with the dispersion of 0.068 dex.
Metallicities of our ELSBGs show a flat tendency with 
increased stellar masses compared with SDSS star-forming galaxies.
It suggests that with the same increments in stellar masses, the center region gas-phase metallicity in LSBGs 
are not enhanced as much as in normal galaxies.
The gas-phase metallicity enhancement in LSBGs is inefficient.

\begin{figure}
  \begin{center}
  \includegraphics[width=3.0in]{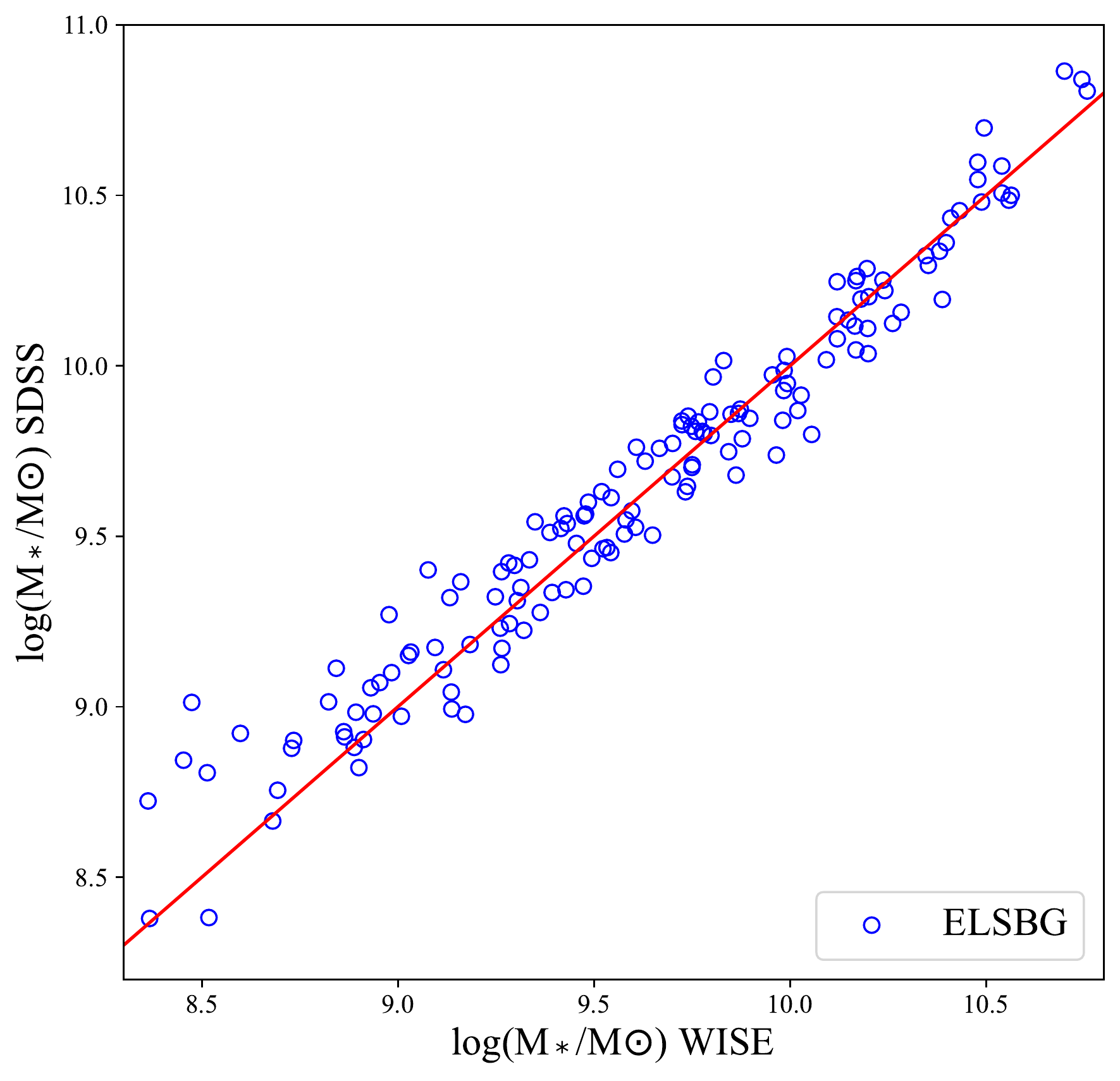}
  \end{center}
\caption{The open blue circles are 153 ELSBGs with S/N of W2 band larger than ten and the number of blends equaling one. 
The red line represent the stellar mass derived from SDSS equaling to the stellar mass derived from WISE.}
       \label{mass}
\end{figure}

Considering the dust extinction could affect stellar mass derived from optical luminosity for edge-on galaxies \citep{2006A&A...456..941M},
we use Wide-field Infrared Survey Explorer (WISE; \citealt{2010AJ....140.1868W}) data to check the stellar mass.  
We apply the method from \cite{2012AJ....143..139E} to estimate the stellar mass using W1 and W2 bands. 
We match positions of our ELSBGs with the WISE All-Sky Source Catalog from the dataset 10.26131/IRSA142 and constrain the S/N of the W2 band larger than ten and the number of blends 
equaling one. 153 ELSBGs satisfy our criteria. We show the result in Figure\,\ref{mass}. 
Except for a few low-mass ELSBGs, the stellar mass derived from SDSS and WISE is consistent with a dispersion value of 0.095 dex.
Overall, the dust extinction does not affect stellar mass derived from SDSS z-band for our sample.

\subsection{Gas to Stellar Mass Ratio and Chemical Evolution}

The gas-to-stellar mass ratio is an important clue to the conversion from the gas into stars. 
The Arecibo Legacy Fast Arecibo L-band Feed Array (ALFALFA, \citealt{2005AJ....130.2598G}) 
performs a very wide area (7000 deg$^2$) blind extragalactic {\ion{H}{I}} survey.
We match our 330 blue ELSBGs position information with the optical positions from 
 the full ALFALFA ($\alpha$.100) catalog \citep{2020AJ....160..271D}. The matching radius is fixed in 3$''$.
43\% of our sample is uncovered by ALFALFA and 34\% is unmatched (undetected) with ALFALFA.
There are 75 ELSBGs matched with ALFALFA, and the other two ELSBGs (UGC 07301 and UGCA 014) have {\ion{H}{I}} data 
from \cite{2005ApJS..160..149S} and \cite{1988ApJS...67....1T}.
We estimate the total gas mass by M$\rm _{gas}$ = 1.36 $\times$ M$\rm_{HI}$ 
\citep{2006ApJ...647..970L, 2017ApJ...837..152D}.
The gas fraction is derived by f$\rm _g$ = M$\rm _{gas}$/(M$\rm _{gas}$+M$_*$). 
f$\rm _g$ ranges from 97\% to 61\% with a mean value of 84\% for our blue ELSBGs. 

The closed-box model of chemical evolution predicts the metallicity by 
mass (\citealt{1963ApJ...137..758S}; \citealt{1972ApJ...173...25S}): Z = y ln $\rm f_g^{-1}$, 
where yield value (y) represents the ratio of the
mass of newly formed metals to the mass of gas permanently
locked into stars. This relation can be written as \citep{2004MNRAS.355..887K}:
12 + log(O/H) = 12 + log(0.196y$_o$) + log\,log\,(1/f$\rm _g$) 
where y$_o$ is oxygen yield and the closed-box model predicts a slope of unity.
Deviations from the slope indicate either infall or outflow of gas.

\begin{figure}
  \begin{center}
  \includegraphics[width=3.8in]{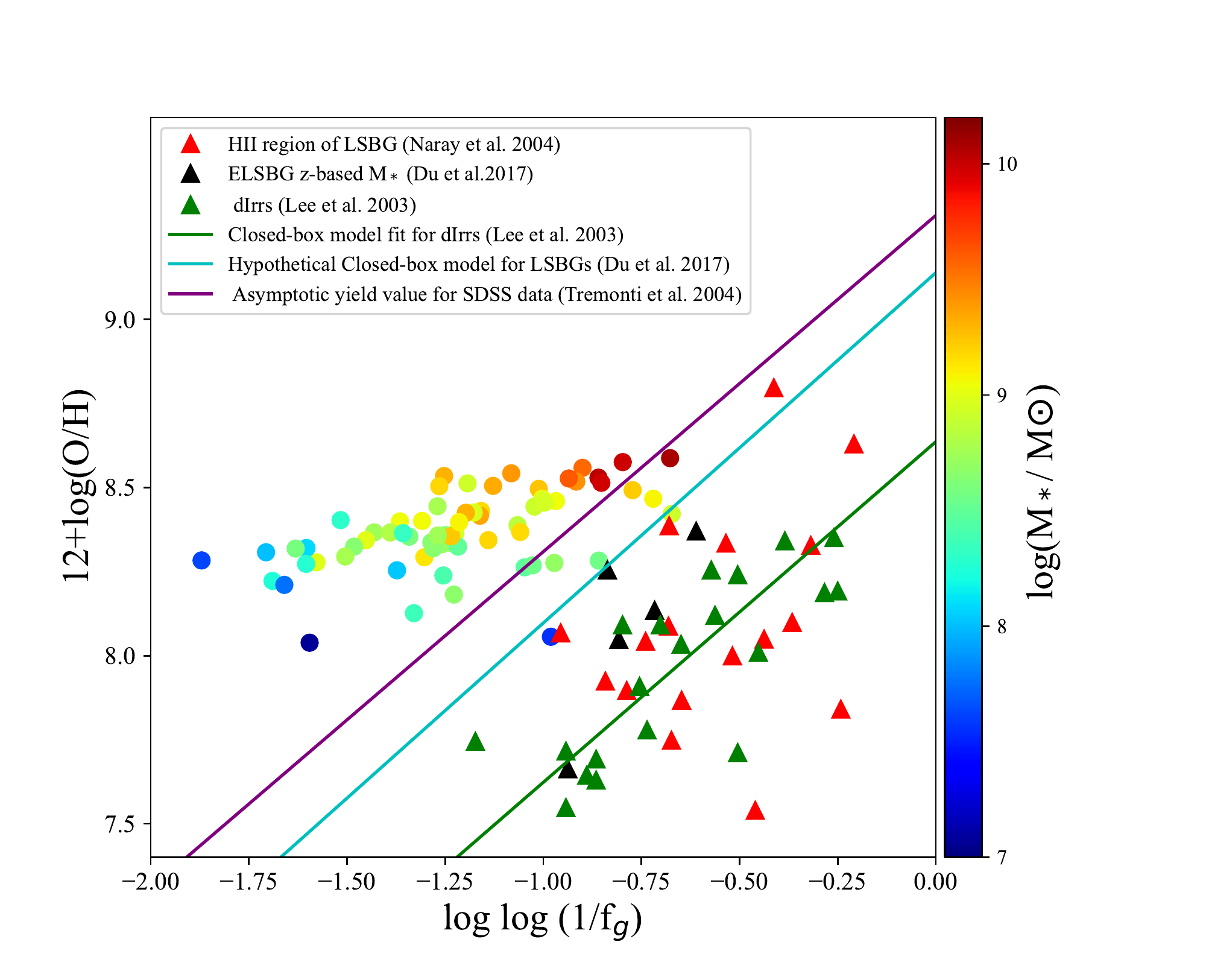}
  \end{center}
\caption{Gas-phase metallicity versus gas mass fraction (log\,log\,(1/f$\rm _g$)) for LSBGs. 
Circle represents our blue ELSBGs. Red and black triangles are LSBGs from \citet{2017ApJ...837..152D}
and \citet{2004MNRAS.355..887K}, respectively. Green triangles are dIrrs from \cite{2003AJ....125..146L}.
The green solid line with a slope of unity represents a closed-box model of chemical evolution with y$_o$ = 2.22$\times$10$^{-3}$ for dIrrs sample from 
\cite{2003AJ....125..146L}.
The cyan solid line is a `hypothetical' closed-box model for LSBGs in \citet{2017ApJ...837..152D}
 with an oxygen yield 6.23$\times$10$^{-3}$. 
 The purple line is the asymptotic yield value for SDSS data \citep{2004ApJ...613..898T} with y$_o$ = 0.0104.
 The color code shows the stellar mass of our blue ELSBGs.}
           \label{gas}
\end{figure}

Figure\,\ref{gas} shows the gas-phase metallicity versus gas mass fraction (log\,log\,(1/f$\rm _g$)).
\citet{2017ApJ...837..152D} discussed that z-band based M$_*$ from MPA-JHU catalog are realiable for ELSBGs
and their f$\rm _g$ based on z-band M$_*$ ranges from 76\% to 57\% in Figure\,\ref{gas}.
The stellar masses of our blue ELSBGs are also based on z-band from the MPA-JHU catalog.
The f$\rm _g$ based on B-band M$_*$ of LSBGs in \citet{2004MNRAS.355..887K} ranges from 77\% to 23\% with the mean value of 50\%. 
ELSBGs from \cite{2017ApJ...837..152D} and LSBGs from \citet{2004MNRAS.355..887K}
follow the closed-box model fit for dIrrs with a large scatter in Figure \,\ref{gas}.
Metallicities of \cite{2003AJ....125..146L} are also measured from {\ion{H}{II}} regions.
Our blue ELSBGs have a higher gas fraction and deviate from the closed-box model fit for {\ion{H}{II}} regions.

SDSS star-forming galaxies in Figure\,\ref{metal} do not follow the closed-box model 
and via galactic winds lossing their metals \citep{2004ApJ...613..898T}. 
Combine with the result of Figure\,\ref{metal},
the closed-box model can not well explain the ($M_*-Z$) relation for our blue ELSBGs.
\cite{2002AJ....124.1360G} proposed that the continuing gas infall dilutes the metallicity of
those gas-rich and disk-dominated LSBGs.
Besides gas infall, metal-enriched hot gas ($\sim$10$^6$ K) outflow is a key point 
to regulate the mass-metallicity relation \citep{2012ApJ...754...98B, 2019ApJ...886...74M}.
There may be a more complicated chemical evolution pathway with outflow or/and infall for LSBGs 
as the discussion of \cite{2017ApJ...837..152D}.

\subsection{Stellar Populations of Normal ELSBGs and Massive ELSBGs}

Analyzing stellar populations is the archaeology of studying galaxies.
Stellar population synthesis is a widely used method to obtain stellar populations 
of galaxies. It requires a high-quality continuum of the spectrum.
The continuum S/N of the single spectrum (S/N $<$ 10) of our ELSBGs 
can not satisfy stellar population synthesis.
We divide our sample into five stellar mass bins as shown in Table\,\,\ref{table:bins} 
and stack the fiber spectra in every stellar mass bin.
Firstly, we shift the single spectrum to the rest frame, then interpolated onto 
a wavelength grid spanning 3700-9000 $\rm \AA$ with a step of 3 $\rm \AA$.
Every spectrum is normalized by 5500-5600 $\rm \AA$ and 
we stack spectra using the mean flux density at each wavelength.
The error of the stacked spectrum is estimated by the standard deviation of data points at each wavelength.

We perform the stellar population synthesis by STARLIGHT \citep{2005MNRAS.358..363C}
 to fit the observed spectrum O$\rm _{\lambda}$ with a model spectrum M$\rm _{\lambda}$ from
 BC03 model spectra \citep{2003MNRAS.344.1000B}. 
We apply 45 spectral components with three metallicities (0.004, 0.02, and 0.05)
 and 15 different ages between 1 Myr and 13 Gyr.
The initial mass function (IMF) is adopted \citet{2003PASP..115..763C}
and the redding law is the `Calzetti law'. The emission lines are masked by STARLIGHT. 
We randomly produced 500 spectra of every stellar mass bin by assuming a Gaussian distribution of error in spectrum,
and input 500 spectra into STARLIGHT to obtain the stellar age and stellar metallicity. 
The error of the resulting output from STARLIGHT is estimated by the standard deviation of those 500 data points.

\begin{figure*}
  \begin{center}
  \includegraphics[width=6.5in]{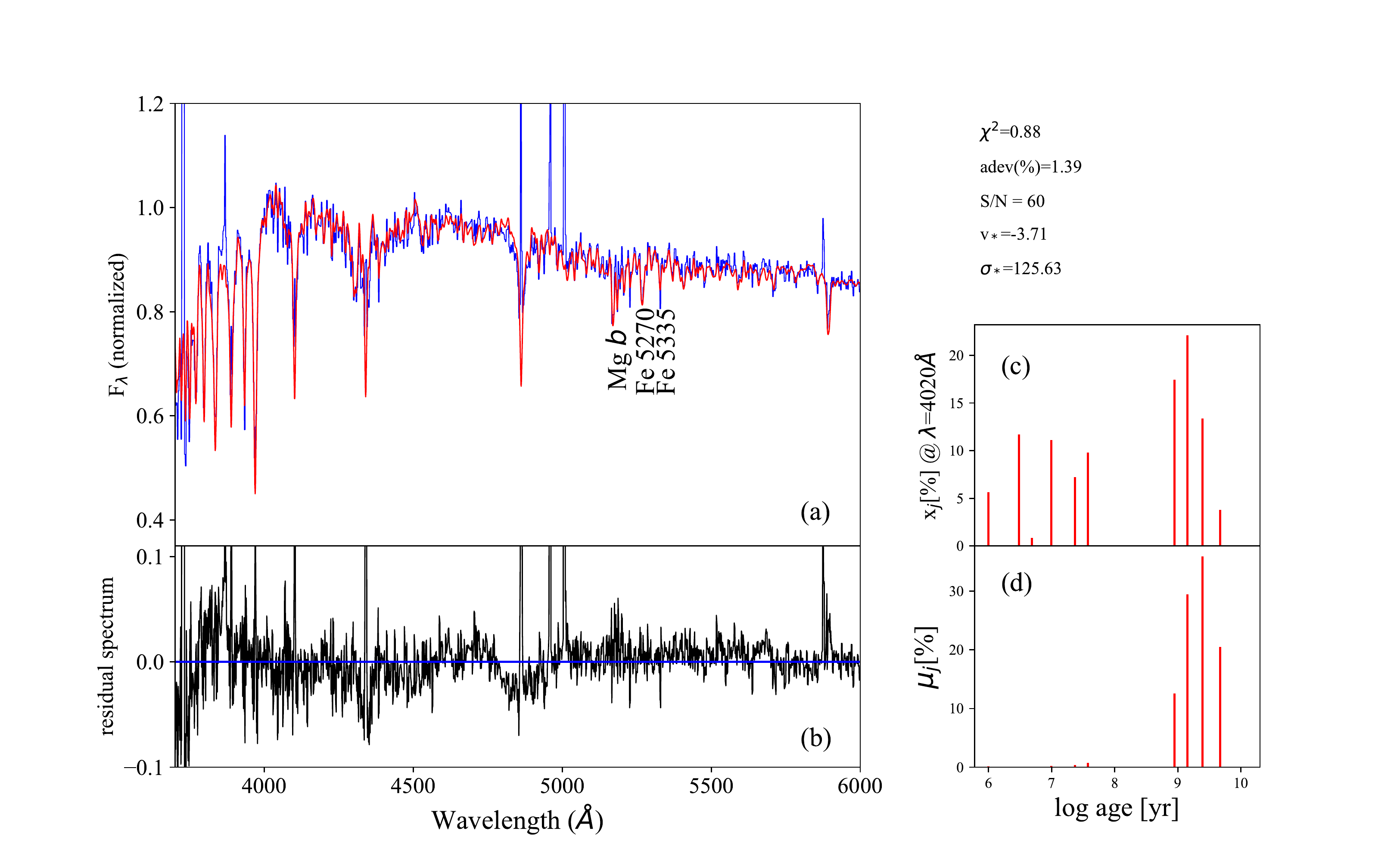}
  \end{center}
\caption{Stellar population synthesis results for the average spectrum of normal ELSBGs using STARLIGHT with 45
SSP from BC03 model spectra.
Panel (a) shows the stacked spectra in the range of 3700-6000 $\rm \AA$ (blue line) and STARLIGHT fit (red line). 
We mark absorption lines: Mg\,$b$, Fe$_{5270}$, and Fe$_{5335}$.
Panel (b) shows the residual spectrum. The blue line is residual flux F$\rm _{res}$ = 0. Panel (c) and panel (d) 
show the light-weighted mean stellar age distributions
and the mass-weighted mean stellar age distributions, respectively.
The five parameters are listed in the top right corner: the reduced $\chi^2$, the mean relative difference
between synthesis and observed spectra $\rm \Delta_{\lambda}$, the S/N in the range of 4730 to 4780 $\rm \AA$, the velocity v$_*$,
 and the velocity dispersion $\rm \sigma_*$.}
             \label{sspm}
\end{figure*}

\begin{figure*}
  \begin{center}
  \includegraphics[width=6.5in]{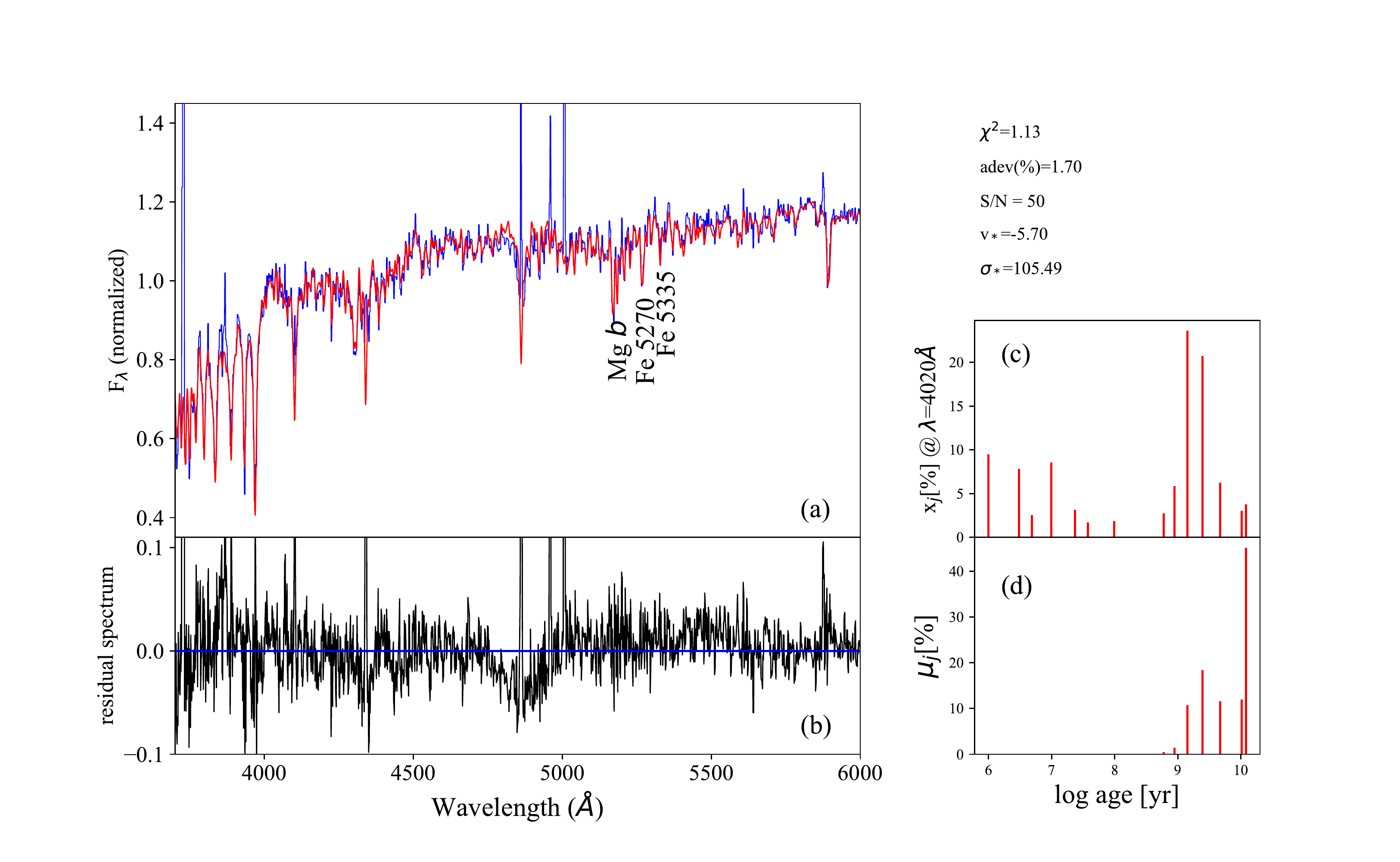}
  \end{center}
\caption{Stellar population synthesis results for the average spectrum of massive ELSBGs using STARLIGHT with 45
SSP from BC03 model spectra.
Panel (a) shows the stacked spectra in the range of 3700-6000 $\rm \AA$ (blue line) and STARLIGHT fit (red line). 
We mark absorption lines: Mg\,$b$, Fe$_{5270}$, and Fe$_{5335}$.
Panel (b) shows the residual spectrum. The blue line is residual flux F$\rm _{res}$ = 0. Panel (c) and panel (d) 
show the light-weighted mean stellar age distributions
and the mass-weighted mean stellar age distributions, respectively.
The five parameters are listed in the top right corner: the reduced $\chi^2$, the mean relative difference
between synthesis and observed spectra $\rm \Delta_{\lambda}$, the S/N in the range of 4730 to 4780 $\rm \AA$, the velocity v$_*$,
 and the velocity dispersion $\rm \sigma_*$.}
             \label{sspl}
\end{figure*}

Table\,\,\ref{table:bins} shows the results of every stellar mass bin. 
\citet{2005MNRAS.358..363C} reported the light-weighted mean stellar age ($<$t$_*$$>$$\rm _L$) 
and light-weighted mean stellar metallicity ($<$Z$_*$$>$$\rm _L$) is more useful in practice.
In the following discussion, we use light-weighted mean stellar age and light-weighted mean stellar metallicity.
We divide the contribution of a light fraction of the stellar population into two age bins, 
young stellar population (age $<$ 1Gyr) and old stellar population (age $>$ 1Gyr).
The stellar mass bin with M$_*$ larger 10$^{9.5}$M$\odot$ show higher $<$t$_*$$>$$\rm _L$ and $<$Z$_*$$>$$\rm _L$ 
than that of low stellar mass bins, and is dominated by old stellar population as shown in Table\,\,\ref{table:bins}.
We divide our sample into massive ELSBGs (M$_*$ $>=$ 10$^{9.5}$M$\odot$) and normal ELSBGs (M$_*$ $<$ 10$^{9.5}$M$\odot$) 
according to their stellar population properties. 
We use the same method as the stellar mass bin to get the stacked spectrum and stellar population properties.
Figure\,\ref{sspm} and Figure\,\ref{sspl} show the results of STARLIGHT fitting for normal ELSBG and massive ELSBG, respectively. 
The results are listed in Table\,\,\ref{table:bins}. Both massive ELSBG and normal ELSBG are metal-poor.
The massive ELSBG show older $<$t$_*$$>$$\rm _L$ ($\sim$2Gyr) than that of normal ELSBGs ($\sim$1Gyr).

\begin{table*}
  \caption{starlight stellar population synthesis results} 
  \centering 
  \begin{tabular}{c c c c c c c} 
  \hline\hline 
  1 & 2 & 3 & 4 & 5 & 6 & 7 \\
  \hline 
  Group & log(M$_*$/M$\odot$) &Number & S/N & $<$log (t$_*$/yr)$>$$\rm _L$ & $<$Z$_*$$>$$\rm _L$ &  YSP$\rm _{LF}$ \\  
  \hline 
  \\
  bin-1 & log(M$_*$/M$\odot$) $>=$ 10.0  & 49 & 45 & 9.30  & 0.0097 & 41.0\%\\ 
    & & &  & $\pm$0.040  &$\pm$0.0009& \\
  bin-2 & 10.0 $>$ log(M$_*$/M$\odot$) $>=$ 9.5 & 63 & 43 & 9.28 & 0.0091 &  45.5\%\\
  & & &  & $\pm$0.041 & $\pm$0.0010& \\
  bin-3 & 9.5 $>$ log(M$_*$/M$\odot$) $>=$ 9.0 & 84 & 54 & 9.07 & 0.0059 &  57.5\% \\
  & & &  & $\pm$0.048 & $\pm$0.0012 & \\
  bin-4 & 9.0 $>$ log(M$_*$/M$\odot$) $>=$ 8.5 & 90 & 35 & 8.96 & 0.0076 &  67.6\%\\
  & & &  & $\pm$0.033 & $\pm$0.0010& \\
  bin-5 &  log(M$_*$/M$\odot$) $<$ 8.5  & 44 & 31 & 8.76 & 0.0067 &  91.8\% \\
  & & &  & $\pm$0.077 & $\pm$0.0017 & \\
  \hline 
  \\
  Massive ELSBG & log(M$_*$/M$\odot$) $>=$ 9.5  & 112 & 50 & 9.31 & 0.0090 &  43.63\% \\
  & & &  & $\pm$0.039 & $\pm$0.0011& \\
  Normal ELSBG & log(M$_*$/M$\odot$) $<$ 9.5  & 218 & 60 & 8.99 & 0.0048 &  63.77\% \\
  & & &  & $\pm$0.040 & $\pm$0.0010& \\
  \hline
  \end{tabular}\\
  \begin{tablenotes} 
    \footnotesize
   \item{Col.1: The Group number;}
   \item{Col.2: Stellar mass rangement of the bin;}
   \item{Col.3: The number of galaxies in the stellar bin;}
   \item{Col.4: The S/N of stacked spectrum at $\lambda$ 4020$\rm \AA$;}
   \item{Col.5: The light-weighted mean stellar age $<$log (t$_*$/yr)$>$$\rm _L$ from STARLIGHT fitting;}
   \item{Col.6: The light-weighted mean stellar metallicity $<$Z$_*$$>$$\rm _L$ from STARLIGHT fitting;}
   \item{Col.7: The light fraction of young stellar population from STARLIGHT fitting.}
  \end{tablenotes}
  \label{table:bins} 
  \end{table*}

\begin{figure*}
  \begin{center}
  \includegraphics[width=6.5in]{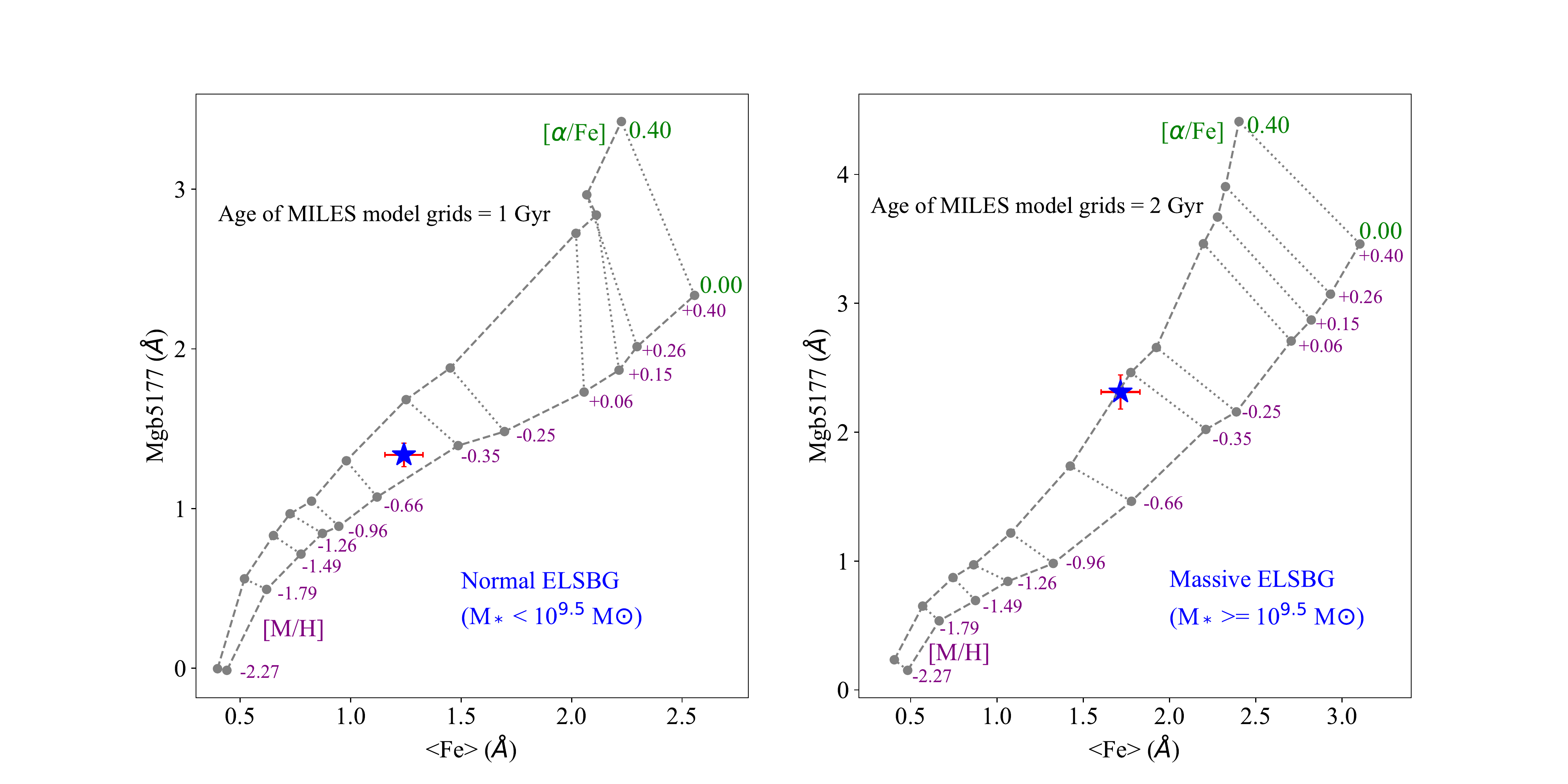}
  \end{center}
\caption{The Mg\,$b$ versus $<$Fe$>$. The left panel shows grids of MILES SSP models fixed at 1Gyr 
for normal ELSBG. The right panel shows grids of MILES SSP models fixed at 2Gyr for massive ELSBG.}  
    \label{MFe}        
\end{figure*}

\subsection{[$\alpha$/Fe] Ratio} 

Magnesium is a good indicator of $\alpha$ elements \citep{2003MNRAS.339..897T, 2020ApJ...899L..12R}
and MILES stellar library \citep{2006MNRAS.371..703S} provides variable age, [$\alpha$/Fe], and [M/H].
We use absorption lines strengths of Mg\,$b$, Fe$_{5270}$, and Fe$_{5335}$ to obtain the [$\alpha$/Fe] ratio 
by comparing single stellar population (SSP) grids from MILES stellar library against the stacked spectrum.
The index definitions of those absorption lines are from \cite{1997ApJS..111..377W}.
SSP grids in low metallicity regions are very narrow and require a high S/N of the continuum, so we use 
the stacked spectrum of massive ELSBGs and normal ELSBGs to get mean [$\alpha$/Fe] ratio. 
We note that the mean [$\alpha$/Fe] ratio represents the value in center region.

We plot the Mg\,$b$ and $<$Fe$>$ = (Fe$\rm _{5270}$+Fe$\rm _{5335}$)\,/\,2 onto the SSP 
grids from MILES stellar library, which has similar spectra resolution to SDSS spectra.
The age of MILES SSP model grids is fixed at 1 Gyr for normal ELSBGs, and 2 Gyr for massive ELSBG which is close to 
the light-weighted mean stellar age from STARLIGHT.
The [M/H] $\approx$ -0.51 (Z$_*$ $\approx$ 0.006) for normal ELSBG
and $\approx$ -0.41 (Z$_*$ $\approx$ 0.0072) for massive ELSBG from interpolation of the selected model.
The Z$_*$ is consistent with the result from STARLIGHT for normal ELSBG and massive ELSBG within 3$\sigma$ error.
From normal ELSBG to massive ELSBG, the mean [$\alpha$/Fe] ratio increases from 0.18 to 0.4 in Figure\,\ref{MFe}.
It is consistent with the result from \cite{2010A&A...520A..69Z} that Mg$_2$ increases with the increasing stellar mass of LSBGs,
because Mg$_2$ increases with increasing [$\alpha$/Fe] ratio.
The [$\alpha$/Fe] ratio is roughly constant ($\sim$ 0.25) over the whole stellar mass range from simulation result \citep{2016MNRAS.461L.102S}.

Several scenarios can explain the results in Figure\,\ref{MFe}:
(1) The shape of IMF. A top-heavy IMF can cause a higher fraction of {\ion{SNe}{II}} compared to SNe\, Ia 
and lead to a higher $\alpha$-enhancement \citep{2017MNRAS.464.3812F}.
The high star formation rate environments usually have top-heavy IMFs \citep{2010ApJ...708..834B, 2023Natur.613..460L}.
However, LSBGs are low SFR environments. (2) AGN feedback in massive galaxies (M$_*$$>$10$^{10.5}$M$\odot$) 
could enhance the $\alpha$ elements \citep{2016MNRAS.461L.102S}.
But, there is no presence of AGN in our sample.
(3) The time scale of star formation. The $\alpha$-enhancement is mainly related to 
a short-time star-formation activity within 1Gyr because the timescale for SNe\,Ia 
(produce iron-peak elements diluting $\alpha$ elements) is about 1Gyr.
The slow star-formation process in LSBGs may prolong the $\alpha$-enhancement timescale.
Furthermore, low stellar surface density in LSBGs also could make a lower probability of 
SNe\,Ia explosion than that in normal galaxies.
We can observe $\alpha$-enhancement with a longer timescale in LSBGs.
(4) Metal-rich gas outflow event caused by SNe\,Ia winds which can take away the iron-peak elements.
The simulation results show that the supernovae' energy injection rate (that goes into the wind) 
and the outflow rate is inversely correlated with the gas surface density 
($\Sigma_{\rm gas}$) \citep{2017MNRAS.470L..39F}.
The mean $\Sigma_{\rm gas}$ of LSBGs is about 4.1 M$\odot$ pc$^{-2}$ and 
lower than that of star-forming galaxies \citep{2019ApJS..242...11L}. 
The effect of supernovae wind in LSBGs is stronger than that in star-forming galaxies. 
This scenario also supports the discussion in Section 3.2.

However, the study on the supernovae in LSBGs is very rare \citep{2012A&A...538A..30Z}.
More observational evidence is needed to corroborate our discussions.
Due to the spectroscopic data being extremely limited for LSBGs,
in the future, integral field spectroscopy is vital to follow up on our 
results.

\section{Conclusions}
In this paper, we present a sample of 330 blue ELSBGs. 
Our sample is selected by color-mass relation.
We discussed the chemical evolution of our sample.
Our main findings are as follows:

1. We derived the gas-phase oxygen abundance by [{\ion{N}{II}}]$\lambda$6583/H$\alpha$ diagnostic. 
In the gas-phase ($M_*-Z$) relation plot, 
metallicities of our ELSBGs show a flat tendency compared with SDSS star-forming galaxies, indicating the oxygen
abundance enhancement is inefficient in LSBGs. 

2. 77 ELSBGs have {\ion{H}{I}} data and are gas-dominant galaxies. 
We analyze that the chemical evolution of our ELSBGs can not be explained by the closed-box model
according to the gas-phase ($M_*-Z$) Relation plot and the relationship between metallicity and gas mass fraction (log\,log(1/f$\rm _g$)). 
The complicated chemical evolution model (including metal-poor gas infall and/or metal-rich gas outflow) 
may exist in LSBGs.

3. We separated our blue ELSBGs into normal ELSBGs (M$_*$$<$10$^{9.5}$M$\odot$) and 
massive ELSBGs (M$_*$$>=$10$^{9.5}$M$\odot$) according to their stellar population 
properties. Both normal ELSBGs and massive ELSBGs show lower $<$Z$_*$$>$$\rm _L$ than solar metallicity. 
The $<$t$_*$$>$$\rm _L$ of normal ELSBG and massive ELSBG are 0.97\,Gyr and 2.04\,Gyr, respectively. 
Massive ELSBG is dominated by old stellar populations ($>$1Gyr) 
and normal ELSBG is dominated by young stellar populations ($<$1Gyr).

4. We obtain the mean [$\alpha$/Fe] ratio by the line strengths of Mg\,$b$, and $<$Fe$>$ through the MILES SSP grids.
From normal ELSBG to massive ELSBG, the mean [$\alpha$/Fe] ratio increases from 0.18 to 0.4.
The time scale of star-formation and/or metal-rich gas outflow event caused by SNe\,Ia winds, 
have been considered to be responsible for the $\alpha$-enhancement of ELSBGs.

\section*{Acknowledgments}
We thank an anonymous referee for a number of very constructive comments.
We thank Dr.\,Jia-Sheng Huang, and Dr.\,Wei Zhang for
their insightful comments and/or useful communications
during the preparation of the manuscript.
H.\,W. acknowledges the National Natural Science Foundation of China (NSFC) grant No.\,11733006 and 12090041.
G.\,G. acknowledges the ANID BASAL projects ACE210002 and FB210003. 
C.\,C. is supported by the NSFC, 
No. 11803044, 11933003, 12173045 and acknowledge the science research grants 
from the China Manned Space Project with NO. CMS-CSST-2021-A05.
J.\,W. acknowledges the NSFC grants U1831205, 12033004, 12221003 
and the science research grants from the China Manned Space Project 
with NO. CMS-CSST-2021-A06 and CMS-CSST-2021-B02.
T.\,C. acknowledges the NSFC grant No. 12173045 and 12073051.
This work is sponsored (in part) by the Chinese Academy of Sciences (CAS), 
through a grant to the CAS South America Center for Astronomy (CASSACA).

We dedicate this work to the remembrance of Dr. Yu Gao, 
a respected scientist passed away in May 2022, who successfully detected molecular gas in ELSBGs.

\bibliographystyle{apj}
\bibstyle{thesisstyle}
\bibliography{main}

\end{document}